# Environmental injustice in America: Racial disparities in exposure to air pollution health damages from freight trucking

**Priyank Lathwal,[a] Parth Vaishnav,[b],* M. Granger Morgan[c]**

[a] Belfer Center for Science and International Affairs, Harvard University, Cambridge, MA 02138

[b] School for Environment and Sustainability, University of Michigan, Ann Arbor, MI 48109

[c] Department of Engineering and Public Policy, Carnegie Mellon University, Pittsburgh, PA 15213

*Corresponding Author; Email: parthv@umich.edu

**This PDF file includes:**

> **Main Text:** 2,609 words (excluding abstract, methods, figure legends, references)
> **Figures:** 1 to 4
> **Tables:** 1 to 3

**Abstract**

$PM_{2.5}$ produced by freight trucks has adverse impacts on human health. However, it is unknown to what extent freight trucking affects communities of color and the total public health burden arising from the sector. Based on spatially resolved US federal government data, we explore the geographic distribution of freight trucking emissions and demonstrate that Black and Hispanic populations are more likely to be exposed to elevated emissions from freight trucks. Our results indicate that freight trucks contribute ~10% of $NO_x$ and ~12% of $CO_2$ emissions from all sources in the continental US. The annual costs to human health and the environment due to $NO_x$, $PM_{2.5}$, $SO_2$, and $CO_2$ from freight trucking in the US are estimated respectively to be $11B, $5.5B, $110M, and $30B. Overall, the sector is responsible for nearly two-fifths (~$47B out of $120B) of all transportation-related public health damages.

**Main**

Transportation is the largest contributor to US greenhouse gas (GHG) emissions, accounting for 27% of all GHG emissions by sector in 2020[1] and contributing to roughly 17,000 to 20,000 premature deaths each year[2]. Within the transportation sector, medium and heavy-duty vehicles (MHDVs), make up for nearly a quarter of all US transportation greenhouse gas (GHG) emissions[1]. These vehicles are the largest contributor of oxides of nitrogen ($NO_x$) along with other pollutants such as fine particulate matter ($PM_{2.5}$), and GHGs, especially carbon dioxide ($CO_2$)[3]. $PM_{2.5}$ from heavy duty vehicles has adverse health impacts such as heart disease, lung cancer, and respiratory illnesses[3–5] on the ~72 million people that reside near freight truck routes and other congested areas in America[6]. Many of these tend to be low-income communities and people of color that have been exposed to elevated levels of truck pollution for decades[7]. As other transportation modes such as passenger vehicles become cleaner, and the volume of freight truck VMT[8] grows, the proportion of emissions from freight trucking will likely increase in the coming decades. Therefore, it is crucial to reduce the emissions and public health burden arising from US freight trucks.

**Environmental regulation to curtail freight trucking air pollution**

Over the last four decades, air pollutant emissions have declined steadily in the US due to federal environmental regulation such as the Clean Air Act (CAA) [9], and the switch to ultralow sulfur fuel diesel (ULSD) [10]. Much of this decrease has been possible due to putting in place vehicle emissions standards and promoting fuel efficiency. More recently, the Biden-Harris administration has announced a slew of measures[11] as part of their initiative on Strengthening American Leadership in Clean Cars and Trucks[12]. Included in these actions is a newly proposed rule[13] to reduce $NO_x$ emissions from trucks by ~60% by 2045[6] and set stringent GHG standards for certain heavy-duty vehicle categories. Together, these actions are directed towards reducing the emission burden from freight trucks and attempt to provide benefits to communities of color and low-income communities[11]. However, despite being the fastest sector in terms of emissions growth, decarbonization of heavy-duty trucking has lagged compared to other sectors[14].

**Absolute emissions have declined but relative disparities persist**

Although $PM_{2.5}$ concentrations have reduced by as much as 70% since the early 1980s [15], racial-ethnic and socio-economic disparities [16–20] continue to exist [15,21]. While many historical and current policy regimes prioritize emissions reduction, primarily in the form of reducing attributable $PM_{2.5}$ mortality, they provide little guidance on *where* we should focus on advancing environmental justice when implementing air pollution reductions and *how* to address distributional impacts on communities [22]. A recent paper by Tessum et al.[21] found higher than average $PM_{2.5}$ exposures across race and ethnicities in comparison to the white population from different sources. The authors use "population-weighted ambient concentrations" as a unit of measure to report overall disparity and advocate for targeting local emission hotspots as a mitigation strategy to alleviate prevailing systemic $PM_{2.5}$ exposure disparity. Initiatives such as the "Justice40 Initiative" attempt to fix this by providing interim implementation guidance to specific federal agencies and securing environmental justice for disadvantaged communities (DACs).

Yet, when it comes to meeting environmental justice goals in the context of heavy-duty trucking, there exists there is little research and supporting data to inform and improve environmental justice outcomes. In fact, no current study exists that has looked at the environmental, climate, and public health burden of freight trucks on racial and ethnic groups based on a bottom-up truck activity data. Our analysis fills this gap. Based on a bottom-up national truck activity data, this paper conducts a first order assessment by focusing on the environmental, climate, and public health impacts of disparities in air pollution exposure arising from freight trucks in the US. We focus specifically on the freight trucking sector because it is crucial in terms of economic value added to the economy (roughly 72% of the total $10.4T domestic freight was carried by freight trucks in 2017[23]) and has an important role to play for the United States to reach its net zero target by 2050[24].

Through this study, we make three contributions to the literature. First, we conduct a spatially resolved bottom-up emissions assessment, which is based on the most updated available national freight data from the Federal Highway Administration (FHWA): Highway Performance Monitoring System (HPMS) link-level activity data included in Freight Analysis Framework Version 4 (FAF4) [25]. Second, we report environmental, climate, and air pollution related monetized public health impacts due to freight trucking at the county level for the contiguous states. Further, we quantify the extent of air pollution related public health damages that are being exported from or imported to individual counties due to trucking in the US. At any given location, emissions from distances as great as 800 km can cause air pollution related health damages [26]. Therefore, a large fraction of the human health burden within a county or census tract may be due to emissions from trucking that are produced elsewhere. We develop our estimates using publicly available data from the US federal government which we combine with reduced complexity integrated air quality models (RCMs): Estimating Air Pollution Social Impact Using Regression (EASIUR) [27,28] and the source-receptor model Air Pollution Social Cost Accounting (APSCA) [26]. By doing so, we explore the spatial heterogeneity in air pollution damages at the county level based on source-receptor relationships. We focus on air pollution public health damages attributable to outdoor $PM_{2.5}$ exposure because it is responsible for ~90% of all air pollution related health damages [29]. Third, we perform the demographic analysis at the resolution of individual counties and census tracts to evaluate air pollution related health damages and distributional effects. Furthermore, we observe air pollution related disparities across racial and ethnic groups both at the county and at the census tract resolution. **Figure 1** illustrates the method we use to compute freight trucking emissions, public health and climate change impacts and distributional effects on disadvantaged groups.

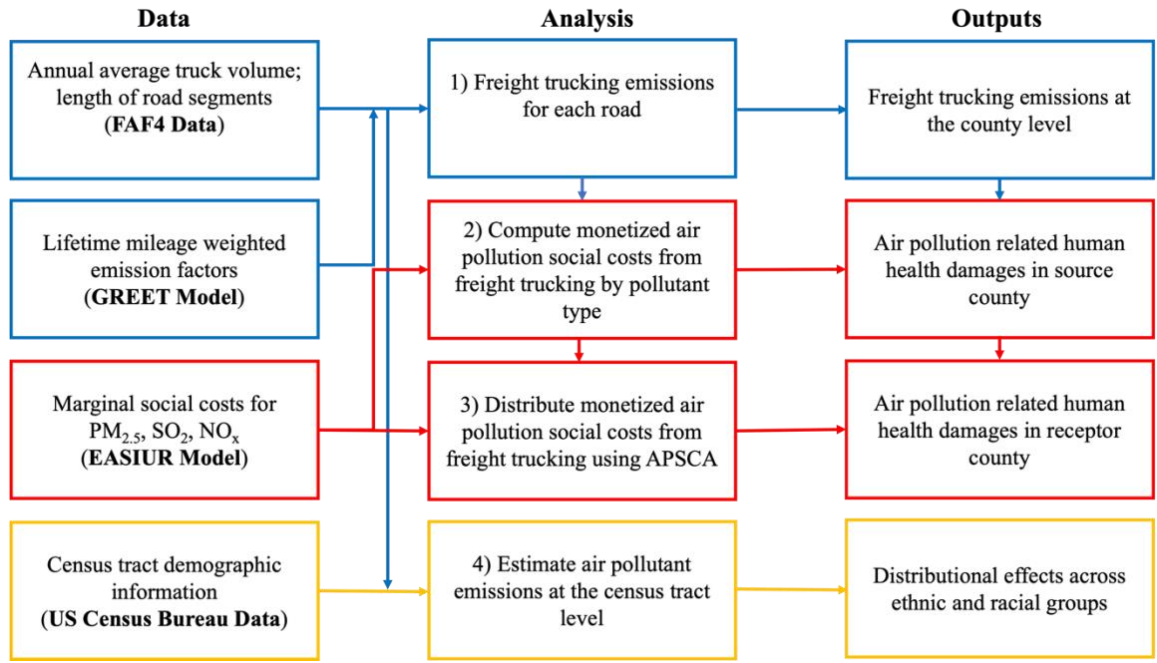

Figure 1. Step process detailing our modeling approach.

**Road freight emissions burden attributable to freight trucks in the US**

**Table 1** shows emissions from freight trucks for the year 2017. These emission estimates are derived from HPMS link-level activity data. We observe that $NO_x$ and $CO_2$ emissions from diesel freight trucks are ~10% and ~12% of the total US emissions. As a sanity check, we also compare these estimates with economy-wide estimates of emissions from the US EPA's 2017 NEI. We direct the reader to **Supplementary Note 1** and **Supplementary Note 2** for comparison results and potential reasons as to why we observe differences in our emission estimates and those obtained from the most recent and publicly available US EPA's 2017 NEI.

**Table 1. Freight trucking emissions estimates derived from HPMS estimates in FAF4 for 2017. The estimate in the last column provides percentage contribution of freight trucking emissions derived from HPMS data (included in FAF4) to total US emissions. This is obtained by dividing the freight trucking emissions derived from HPMS activity data by total emissions from all sources in NEI, 2017.**

| Pollutant/GHG | Trucking emissions derived from HPMS data (tons) | Total NEI emissions (tons) | % US emissions from freight trucking based on HPMS data |
|---|---|---|---|
| $PM_{2.5}$ | 28K | 5.2M | 0.53% |
| $SO_2$ | 4.6K | 2.5M | 0.18% |
| $NO_x$ | 1.1M | 11M | 10% |
| $CO_2$ | 640M | 5.3B | 12% |

**Environmental and Public Health Impacts of Freight Trucking in the US**
We observe that the emissions burden is highest in the counties that include the road network (see **Figure *2*(A)**). **Figure *2*(B)** shows county level spatial distribution of $NO_x$ emissions from freight trucks. $NO_x$ trucking emissions are high in counties in the northeast, southern, and western parts of the US. $PM_{2.5}$ and $NO_x$ emissions from freight trucks are the lowest in Delaware, Vermont, and the District of Columbia.

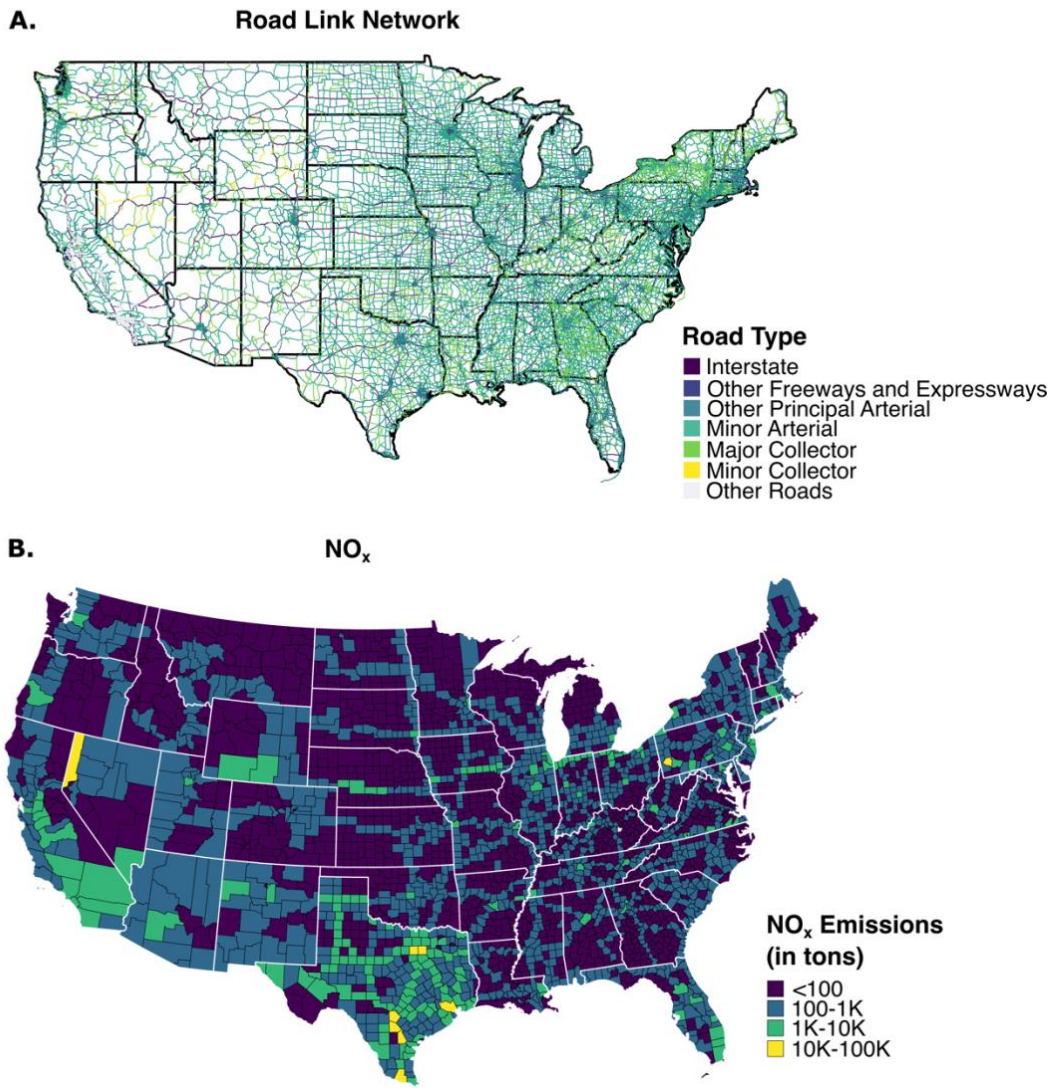

**Figure 2. (A) FAF4 road network in the contiguous US. These routes comprise ~446,000 miles of roads. The dataset includes interstate highways, national highway system (NHS) roads, rural and urban principal arterials along with intermodal connectors. (B) County level spatial distribution of $NO_x$ emissions from MHDV trucking. The spatial distribution of $PM_{2.5}$, and $SO_2$ emissions is similar to $NO_x$ emissions shown in the figure.**

We estimate the total annual public health damage resulting from diesel trucks for $PM_{2.5}$, $SO_2$, and $NO_x$, to be \$5.5B, \$110M, \$11B, respectively (costs expressed in 2017 US dollars), and the societal damage from $CO_2$ emissions to \$30B (in 2017 US dollars), assuming a social cost of carbon of \$51 per ton[30,31]. While the numbers for $PM_{2.5}$, $SO_2$, and $NO_x$, are estimates of the human health damage that occurs everywhere from emissions that originate in each county, here we also separate that estimate into damage that occurs within each county from the emissions within that county, and damage that occurs within the county due to emissions that are "imported" from other counties.

Likewise, we split the estimate of damages that originate in a county into damages that are felt within the same county and damages that are "exported" to other counties. This allows us to estimate the total damage that occurs within each county and whether the county is a net importer or exporter of air pollution damages. We perform this disaggregation using the APSCA model [26] that provides receptor resolved air pollution damages at the county level. **Figure 3(A-C)** show public health $PM_{2.5}$, $SO_2$, and $NO_x$ damages resolved at the source counties due to freight trucks for 2017. This is the total damage from emissions that originate in a particular county and includes the damage caused by source county emissions activity within the county itself. If we exclude the air pollution damage that the source county causes within itself, we observe that the counties located in the states of Texas ($3.4B), Pennsylvania ($1.5B), Indiana ($1.1B), New Jersey ($1B), and New York ($800M) contribute ~49% of all exported air pollution related damages occurring in the US. Even though they are the top 5 exporters of pollution damages, these states are 2nd, 5th, 17th, 11th, and 4th, respectively, by population[32]. The states are 2nd, 6th, 17th, 9th, and 3rd, respectively by gross domestic product[33]. As such, the damages do not correlate precisely with the population and level of economic activity in the states. Instead, damages are determined by the layout of the interstate network, the prevailing winds, and the proximity of states to more populous neighbors.

**Figure *3***(D-F) show public health $PM_{2.5}$, $SO_2$, and NOx damages resolved at the receptor counties due to freight trucks for 2017. This is the total damage a county receives from emissions occurring in other counties, including from the emissions that occur within the county. Cumulatively, the counties located in the states of Texas ($2.7B), New York ($1.4B), Pennsylvania ($1.1B), New Jersey ($930M), and Illinois ($850M) receive ~44% of all imported annual air pollution damage due to freight trucking in the contiguous US.

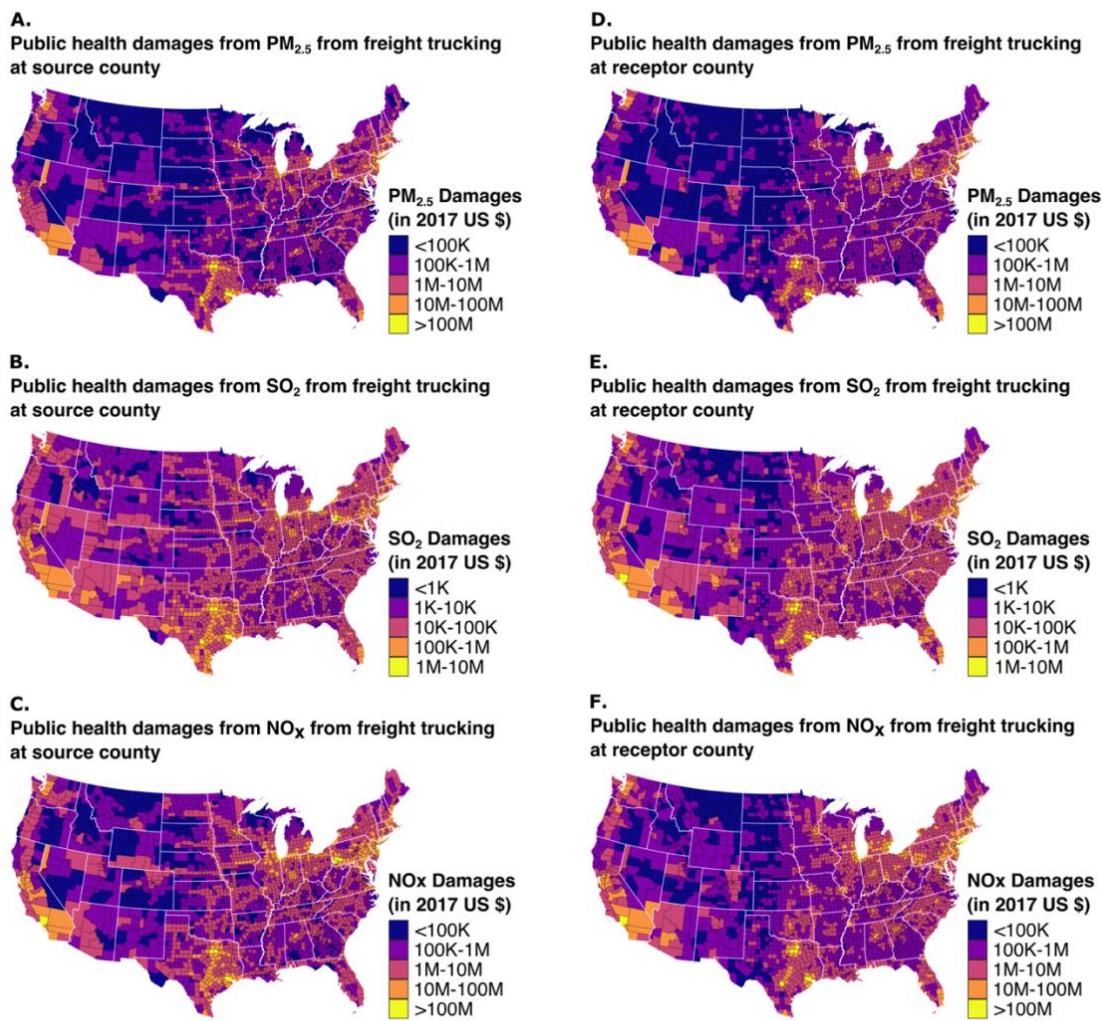

**Figure 3. (A-C) Public health damages from PM$_{2.5}$, SO$_2$, and NO$_x$ due to freight trucks aggregated at the source counties. The source county damage shows total air pollution damage that occurs across all other counties due to the freight trucking activity originating from the source county. It includes the air pollution damage that occurs from the county within itself. (D-F) Public health damages from PM$_{2.5}$, SO$_2$, and NO$_x$ due to freight trucks aggregated at the receptor counties. The receptor damage at a county includes total air pollution damage due to freight trucking activity occurring in other counties. It includes the damage that occurs within the receptor county due to freight trucking activity within the receptor county.**

**Figure *4*** provides map of counties that are net exporters and net importers of freight trucking air pollution related human health damages in the contiguous US. To test the robustness of our results, we do a sensitivity analysis by running the model using another RCM called AP3 and check the results by assessing public health damages from freight

trucking emissions included in NEI, 2017. The results of these comparisons are included in **Supplementary Note 4.2**.

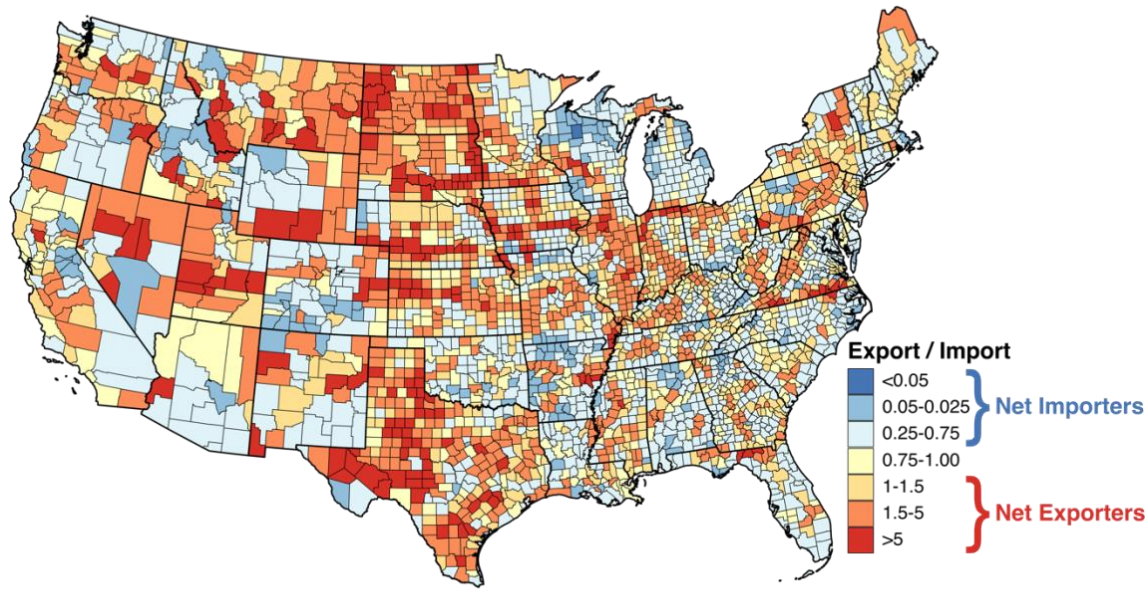

**Figure 4. Map showing counties that are net exporters and importers of freight trucking air pollution related public health damages. We expressed the damages as a ratio by dividing the net exported and imported damage in each county. Damages are expressed in 2017 US dollars.**

**Air Pollution Racial Disparity and Environmental Justice Implications**
The results included below provide a first order estimate of the relationship between air pollution emissions and the proportion of the population each county and census tract that belongs to a disadvantaged subgroup. Our demographic analysis does not allow for any causal assertions as to the reason for the disparity in damages. The location of high levels of freight trucking pollution, and the proximity of highways to disadvantaged racial ethnic subgroups, could be due to the historical practice of redlining and construction of highways through Black and other minoritized communities [34,35]. We observe that $PM_{2.5}$, $SO_2$, and $NO_x$ emissions from freight trucks are significantly higher in counties with higher Black and Hispanic populations (see **Table 2**). Each 1 percentage point increase in the Black population in the county is associated with a 2.1 percentage point increase in $PM_{2.5}$ emissions. Similarly, a 1 percentage point increase in the Hispanic population in the county is associated with a 20-percentage point increase in $PM_{2.5}$ emissions. We also test for this effect in the census tracts where the roads are located and find that the effect is preserved at the census tract level (see **Table 3**). We also observe a negative association between air pollutant emissions and Asian population both at the county and census tract level.

Although a first order analysis, we test the robustness of our results by running the analysis at the county and census tract level on data available from US EPA's 2017 NEI. Both for counties and census tracts, we observe that our findings were consistent with the results obtained by running the analysis on NEI 2017 (see **Supplementary Information Table S7** and **Table S8**). We include a detailed description of the analysis, including the dependent and independent variables in **Supplementary Information**. Additionally, we also run a model specification to gauge whether a county is likely to be an exporter or importer based on the total air pollution damage that the county imports or exports. We find that the US EPA's 2017 NEI showed that counties with larger Black and Hispanic populations were more likely to be net importers of pollution damages (see **Supplementary Information Table S9**Error! Reference source not found.) whereas this conclusion was not supported by emissions estimates derived using our method. This discrepancy in findings is an interesting topic for further investigation.

**Table 2. Effect of freight trucking PM$_{2.5}$, SO$_2$, and NO$_x$ emissions derived from FAF4 on racial and ethnic subgroups at the county level. The dependent variable is the log of PM$_{2.5}$, SO$_2$, and NO$_x$ emissions emitted by freight trucks in counties. For predictor variables that are log values, the relationship can be estimated as %ΔY$_p$ = %β×ΔX$_c$. For predictors that are not log values, the relationship is estimated as %ΔY$_p$ = 100 × (e$^β$-1). These numbers are statistically significant and the numbers in the parenthesis provide standard errors.**

| | Dependent variable: | | |
|---|---|---|---|
| | log(PM2.5) | log(SO2) | log(NOx) |
| | (1) | (2) | (3) |
| Black | 1.120*** | 1.126*** | 1.161*** |
| | (0.159) | (0.163) | (0.187) |
| Asian | -4.849*** | -4.986*** | -5.660*** |
| | (1.305) | (1.338) | (1.538) |
| Hispanic | 3.027*** | 3.045*** | 3.147*** |
| | (0.167) | (0.172) | (0.197) |
| American Indian/Alaskan Native | -0.227 | -0.243 | -0.310 |
| | (0.349) | (0.357) | (0.411) |
| Native Hawaiian/ Other Pacific Islander | -22.638 | -22.323 | -20.848 |
| | (20.179) | (20.682) | (23.768) |
| Other Race | -66.161*** | -68.638*** | -82.887*** |
| | (18.290) | (18.747) | (21.544) |
| Two or More Races | -2.807 | -2.670 | -1.880 |
| | (2.550) | (2.614) | (3.004) |

| | log(PM2.5) | log(SO2) | log(NOx) |
|---|---|---|---|
| log(Median Income) | 0.805*** | 0.823*** | 0.927*** |
| | (0.105) | (0.108) | (0.124) |
| log(Total Population) | 0.646*** | 0.649*** | 0.669*** |
| | (0.018) | (0.018) | (0.021) |
| log(County Area) | 0.317*** | 0.320*** | 0.338*** |
| | (0.026) | (0.027) | (0.031) |
| Constant | -21.281*** | -23.401*** | -19.754*** |
| | (1.216) | (1.246) | (1.432) |
| Observations | 3,106 | 3,106 | 3,106 |
| *Note:* | | | *p<0.1; **p<0.05; ***p<0.01 |

**Table 3. Effect of freight trucking PM$_{2.5}$, SO$_2$, and NO$_x$ emissions derived from FAF4 on racial and ethnic subgroups at the census tract level. The dependent variable is the log of PM$_{2.5}$, SO$_2$, and NO$_x$ emissions emitted by freight trucks in census tracts. For predictor variables that are log values, the relationship can be estimated as %$\Delta Y_p$ = %$\beta \times \Delta X_t$. For predictors that are not log values, the relationship is estimated as %$\Delta Y_p$ = $100 \times (e^\beta - 1)$. These numbers are statistically significant and the numbers in the parenthesis provide standard errors.**

| | *Dependent variable:* | | |
|---|---|---|---|
| | log(PM2.5) | log(SO2) | log(NOx) |
| | (1) | (2) | (3) |
| Black | 0.827*** | 0.832*** | 0.868*** |
| | (0.045) | (0.045) | (0.049) |
| Asian | -0.780*** | -0.751*** | -0.594*** |
| | (0.124) | (0.126) | (0.135) |
| Hispanic | 0.445*** | 0.460*** | 0.536*** |
| | (0.046) | (0.047) | (0.050) |
| American Indian/Alaskan Native | -0.766*** | -0.785*** | -0.892*** |
|                                        | (0.183)   | (0.185)   | (0.199)   |
| Native Hawaiian/ Other Pacific Islander | -5.112**  | -5.270**  | -6.246**  |
|                                        | (2.487)   | (2.513)   | (2.707)   |
| Other Race                             | 0.242     | 0.149     | -0.304    |
|                                        | (1.855)   | (1.875)   | (2.019)   |
| Two or More Races                      | -6.856*** | -6.837*** | -6.754*** |
|                                        | (0.777)   | (0.785)   | (0.845)   |
| log(Median Income)                     | -0.017    | -0.020    | -0.044*   |
|                                        | (0.021)   | (0.021)   | (0.023)   |
| log(Total Population)                  | 0.246***  | 0.246***  | 0.246***  |
|                                        | (0.018)   | (0.018)   | (0.019)   |
| log(County Area)                       | 0.480***  | 0.485***  | 0.520***  |
|                                        | (0.005)   | (0.005)   | (0.005)   |
| Constant                               | -12.424*** | -14.342*** | -9.523*** |
|                                        | (0.264)   | (0.267)   | (0.287)   |
| Observations                           | 57,902    | 57,902    | 57,902    |

*Note:* *p<0.1; **p<0.05; ***p<0.01

**Discussion and Conclusions**

Our results point to several important conclusions. First, we find that freight trucks contribute significantly to $NO_x$ and $CO_2$ emissions in the US, and that counties and census tracts having higher proportions of Black and Hispanic populations are more exposed to elevated emissions from freight trucks. Although disproportionate air pollution impacts have been noted before in general terms in research literature, our work provides a quantitative basis of the contribution of the trucking sector to this effect. The reasons explaining this disparity could be the result of years of racially malign infrastructure siting policy [35,36]. Nonetheless, our findings are salient to inter-agency policy debates at the federal, state, and local level around improving environmental justice outcomes by highlighting the role freight trucking can play in reducing air pollution disparities for DACs. Second, our work identifies regions in United States where environmental inequities are particularly acute due to freight trucks with the aim of aiding new developing approaches to tackle distributional consequences of air pollution from freight trucks and remediation of previous harms in these jurisdictions. More generally, our analysis provides a lens that can be used to explore aspects of environmental justice in other contexts and sectors. Given keen interest in the U.S. government to rebuild the national infrastructure in a way that advances environmental justice, [37–39] policymakers at all levels must consider exploring opportunities under the Infrastructure Investment and Jobs Act (IIJA)[40,41] to clean up legacy pollution by investing in projects that redress environmental harm and advance environmental justice in a meaningful way.

**Materials and Methods**

**Study Area and Scope**

Our study includes freight shipments in the 48 contiguous US states (excluding Alaska, Hawaii, Puerto Rico, and other US territories). The reference year used in the study is 2017, as that is the most recent year for which the national emissions inventory (NEI) is available [42]. This allows us to compare our emissions results with the NEI. The approach we use to extract freight trucking emissions from the NEI is discussed in **Supplementary Note 1**.

**Data**

We use the Federal Highway Administration's (FHWA's) HPMS link-level activity data included in FAF4[25], freight flows through 132 domestic zones in the US. Although a more recent version of the Freight Analysis Framework Version 5 (FAF5) is available, it has yet to be updated with information on county level freight shipments. Thus, we use the HPMS data throughout the analysis. FAF4 data includes a shape file of FAF4 zones providing detailed information on the road network (~446,000 miles; see **Figure 2(A)**) and relevant freight attributes such as the annual average daily traffic volumes on road segments, road lengths and route type. To explore the environmental justice implications of air pollution related damages, we use data from the US Census Bureau [43].

**Estimating freight trucking VMT**

*Road length:* The road network data consist of road links (RLs) and average daily long distance and local truck traffic counts for different freight vehicles. These included long-haul trucks, short-haul freight trucks and non-freight vehicles (buses). We estimate the route road length (in miles) for each road link by taking the difference between the starting and ending mile posts reported in the data for each road link.

*Annual average truck counts:* Freight trucking is mostly diesel powered. So, based on Bickford et al.[44], we assume that 98% of all trucks in our freight data are diesel trucks. Furthermore, the reported average daily truck counts include non-cargo freight vehicles such as commuter and transit buses. In 2012, there were 765,000 bus registrations out of ~58 million truck registrations (excluding sport utility vehicles, vans, and other light vehicles) [45,46]. Thus, to exclusively reflect diesel cargo freight trucks in our truck counts and remove the effect of buses and other non-cargo vehicles, we adjust the daily average truck count on each road link by subtracting 1% of total vehicle counts on each link. We express annual MHDV daily truck traffic on each road link as:

**Equation 3.1**

$$MHDV_{i,daily} = AADTT_{i,daily} * DF * TF$$

**Where,**

$MHDV_{i,daily}$ is the daily average MHDV count on a road link $i$ (expressed as volume per day per section of the road)

$AADTT_{daily}$ is the annual average daily truck traffic for long-distance and non-long distance freight trucks on road link $i$ (expressed as volume per day per section of the road)

$DF$ is the diesel fraction to adjust vehicle counts to include only diesel freight trucks and its value is assumed to be 0.98 from literature [44]

$TF$ is the truck fraction to remove the effect of commuter and transit buses from the annual average daily truck traffic counts. Its value is assumed to be ~0.99 based on vehicle registration data.

*Annual Freight Trucking Vehicle Miles Traveled:* We estimate the daily medium and heavy-duty vehicle miles traveled (MHDVMT) for each road link by multiplying $MHDV_{i,daily}$ by the length of the road segment. We annualize the VMT on each road link by multiplying by 365. However, the MHDVMT obtained is for 2012, and the base year assumed for the analysis is 2017. Using annual VMT data [47] provided by the US Department of Transportation (DOT), we estimate compounded annual growth rate (CAGR) increase in MHDVMT between 2012 and 2017. The MHDVMT in 2017 for each road link is then estimated as:

**Equation 3.2**

$$MHDVMT_i = MHDV_{i,daily} * RL_i * Year_{days} * GF_{VMT}$$

**Where,**

$MHDVMT_i$ is the annualized VMT for medium and heavy-duty vehicles in 2017 for each road link $i$

$MHDV_{i,daily}$ is the daily average medium and heavy-duty vehicle count on a road link $i$ (expressed as volume per day per section of the road)

$RL$ is length of the road segment (in miles) for each road link $i$

$Year_{days}$ is 365, the number of days in a year

$GF_{VMT} = (1 + CAGR_{VMT})^{2017-2012}$ is the growth in MHDVMT between 2012 and 2017. $CAGR_{VMT}$ is estimated to be 2% each year based on authors' calculations from freight trucks' VMT data [47] provided by the US DOT.

**Estimating spatially resolved emissions arising from freight trucking**

We estimate spatially resolved emissions at the county level for $PM_{2.5}$, $SO_2$, $NO_x$, and $CO_2$ in each county by multiplying MHDVMT by the lifetime VMT weighted emission factors (in g/mile) from the Greenhouse Gases, Regulated Emissions, and Energy Use in Transportation (GREET) model [48] for all the road segments contained in the county. For $PM_{2.5}$ emissions, we include tire and break wear emissions in addition to primary $PM_{2.5}$ emissions. Absent better data, we use a constant emission factor regardless of whether the road is in an urban or rural area, even though emission factors for each type may be different. To examine the implications of doing this, we perform a sensitivity analysis with emission factors included in Tong et al. [49]. The results are included in **Supplementary Note 3**. We sum the air pollutant and GHG emission estimates from road links within a county to estimate county level total for 2017. The emissions at the county level are estimated as:

**Equation 3.3**

$$E_{k,p} = \sum_{i \in K} MHDVMT_i * EF_p$$

**Where,**

$E_{k,p}$ is the total MHDV emissions (in tons) for pollutant $p$ ($PM_{2.5}$, $SO_2$, $NO_x$, $CO_2$) in each county $k$,

$MHDVMT_i$ is the VMT on road segment $i$ for MHDVs

$EF_p$ is the emission factor (in g/mile) for pollutant $p$ for the freight truck category under consideration

$K$ is the set of all road segments $i$ contained within county $k$. The sum is performed over all the road segments $i$ that are contained within the county $k$

**Estimating public health and climate damages due to freight trucking**

Using state-of-the-art CTMs to estimate the concentration of air pollutants that results from emissions is very computationally intensive. In order to reduce the computational burden for policy analysis, air quality researchers have developed RCMs to estimate monetized air pollution damages. RCMs divide the entire US into a grid of cells and include a set of look up tables of marginal social costs (MSC; in US $ per ton of pollutant emitted) for emissions associated with each grid cell. For our analysis, we assume that trucking emissions are marginal so that the public health damages due to criteria air pollutants (CAPs) are a simple product of total emissions and MSC for a given pollutant species, location, and height. While the MSC for CAPs is sensitive to location and height, the MSC for $CO_2$ does not depend on these factors, and we estimate damages from $CO_2$

emissions by multiplying emissions by a social cost of carbon (which we assume to be $51 per ton $CO_2$; 2020 US $) [30,31]. We use an RCM called EASIUR [27,28] to estimate marginal damages due to primary $PM_{2.5}$, $SO_2$, and $NO_x$. for 148 x 112 cells with each grid cell 36 km x 36 km in size. Since we are interested in on-road freight transport emissions, we use the annual MSC associated with all area sources with a zero-stack height. We ignore seasonal variation. Next, we conduct an overlay analysis and find how much of each county's area lies within the bounds of each grid cell. We repeat this exercise for each grid cell in the contiguous US and distribute the marginal social costs through a weighting factor based on the area of county contained in each grid cell. Finally, we calculate the marginal damages as a product of the MSC in each county multiplied by the emissions of a particular pollutant species at a given location. EASIUR provides MSC assuming a value of statistical life (VSL) of $8.6M (2010 US $). **Supplementary Note 4** explains how we have updated the MSCs to 2017 dollars using income adjustment and population adjustment while accounting for inflation. For $CO_2$, we assume a social cost of carbon (SCC) of $51 per ton of $CO_2$ [30,31]. Mathematically, this can be expressed as:

**Equation 3.4**

$$MD_p = \sum_k MSC_{k,p} * E_{k,p}$$

**Where,**

$MD_p$ is the marginal damage in 2017 dollars for the US for pollutant $p$ ($PM_{2.5}$, $SO_2$, and $NO_x$)

$MSC_{k,p}$ is the marginal social cost in county $k$ for pollutant $p$ (expressed in US $ per ton of pollutant emitted)

$E_{k,p}$ is the emissions for county $k$ for pollutant $p$ (expressed in tons)

EASIUR provides an estimate of all the damages that occur everywhere from the emissions that *originate* within county, regardless of where those damages occur. To understand *where* damages occur, we employ source-receptor (S-R) relationships from the APSCA model to spatially disaggregate social cost of pollutants [26]. As a boundary check, we also calculate air pollution damage estimates from freight trucking at the source counties using another RCM called the Air Pollution Emission Experiments and Policy Version 3 (AP3) [50]. We find reasonable consistency between the results of each model. The results of the comparison of the public health damages from freight trucking are reported in **Supplementary Note 4**.

**Impact of freight trucking pollution across demographic groups**

While the monetized environmental and climate change impacts are felt across different geographical sub-units, here, we focus on the distributional effects of freight trucking emissions on different ethnic and racial subgroups. This is important because the literature shows that, historically, adverse air pollution-related health impacts have been inequitable [16–20]. Further, most policy interventions do not account for distributional welfare. To evaluate differences in the impacts of pollution from freight trucking on different racial groups, we use county level population data from US Census Bureau's 2010 decennial census [43]. Based on county as the unit of spatial aggregation, we focus on seven self-identified racial and ethnic subgroups, selected so that they are mutually exclusive: (1) Black or African American alone, (2) Asian alone, (3) Hispanic or Latino origin by race, (4) American Indian and Alaskan Native alone, (5) Native Hawaiian and Other Pacific Islander alone, (6) some other race alone, and (7) two or more races. These population subgroups add up to the total population of the county or census tract. We hypothesize an association between $PM_{2.5}$, $SO_2$, and $NO_x$ emitted from freight trucks in different counties and local race-ethnicity populations residing in counties such that for a unit increase in the absolute populations of different racial ethnic groups across counties, we should observe a corresponding increase in absolute $PM_{2.5}$, $SO_2$, and $NO_x$ emissions from freight trucks. We provide a description of the dependent and independent variables included in the analysis along with the hypothesized relationship in **Supplementary Note 6.**

We posit a model specification that attempts to find association of emissions for pollutant $p$ ($PM_{2.5}$, $SO_2$, and $NO_x$) in county $c$ as a function of the demographic and other attributes of the census tract where the emissions occur along with an unobserved error term ($\epsilon_{p,c}$). We control for income to disentangle the effects of economic and race-based disparities, the former of which are well established in the literature[21,51,52]. Further, we include the total population since population may correspond to the level of economic and freight activity. Bigger counties might have more roads and thereby trucks may travel more vehicle miles. Therefore, we control for the area of the county. $\beta$ is the modeled coefficient for each corresponding independent variable $X$ relative to White population in county $c$. The model specification is:

**Equation 3.5**

$$Y_{p,c} = \beta_0 + \beta_{black}X_c^{black} + \beta_{asian}X_c^{asian} + \beta_{hisp}X_c^{hisp} + \beta_{amerind}log\left(X_c^{amerind}\right) + \beta_{haw}X_c^{haw} + \beta_{other}X_c^{other} + \beta_{twomore}X_c^{twomore} + \beta_{medinc}\log\left(X_c^{medinc}\right) + \beta_{totpop}\log\left(X_c^{totpop}\right) + \beta_{area}log(X_c^{area}) + \epsilon_{p,c}$$

Where,

$Y_{p,c}$ is the log of freight trucking emissions for pollutant $p$ ($PM_{2.5}$, $SO_2$, and $NO_x$) in county $c$

$X_c^{black}$ is the proportion of the total population that identifies as non-Hispanic Black or African American alone in county $c$

$X_c^{asian}$ is the proportion of the total population that identifies as non-Hispanic Asian alone in county $c$

$X_c^{amerind}$ is the proportion of the total population that identifies as American Indian and Alaska native alone in county $c$

$X_c^{haw}$ is the proportion of the total population that identifies as Hawaiian and other Pacific Islanders alone in county $c$

$X_c^{other}$ is the proportion of the total population that identifies as some other race alone in county $c$

$X_c^{twomore}$ is the proportion of the total population that identifies as two or more races alone in county $c$

$X_c^{medinc}$ is the median income of the house hold in county $c$

$X_c^{totpop}$ is the total population in county $c$

$X_c^{area}$ is the area of the county $c$

We then perform the same analysis at the census tract level. To calculate the emissions that occur within each census tract, we download census tract shapefiles from the U.S. Census Bureau [53]. Using the shapefile of the road network included in FAF4, we estimate the centroid of each road segment. We then use the "rgeos" package in the R programming environment to identify which census tract each road segment centroid is located in. We repeat the calculations described in **Equation 3.3** but sum the emissions that occur along each road segment over all the road segments that fall within a census tract (instead of summing over all the road segments that fall within a county).

Finally, the effect of trucking emissions on racial-ethnic population subgroups in the census tract is evaluated using linear regression. We propose a model specification that attempts to find the relationship between emissions for pollutant $p$ (PM$_{2.5}$, SO$_2$, and NO$_x$) in census tract $t$ as a function of the demographic and other attributes of the census tract where the emissions occur and an unobserved error term ($\epsilon_{p,t}$). $\beta$ is the modeled coefficient for each corresponding independent variable $X$ relative to the White population in census tract $t$. The model specification is:

**Equation 3.6**

$$Y_{p,t}(X) = \beta_0 + \beta_{black}X_t^{black} + \beta_{asian}X_t^{asian} + \beta_{hisp}X_t^{hisp} + \beta_{amerind}log\,(X_t^{amerind}) + \beta_{haw}X_t^{haw}$$
$$+ \beta_{other}X_t^{other} + \beta_{twomore}X_t^{twomore} + \beta_{medinc}log\,(X_t^{medinc}) + \beta_{totpop}log\,(X_t^{totpop})$$
$$+ \beta_{area}log(X_t^{area}) + \epsilon_{p,t}$$

Where,

$Y_{p,t}$ is the log of freight trucking emissions for pollutant $p$ (PM$_{2.5}$, SO$_2$, and NO$_x$) in census tract $t$

$X_t^{black}$ is the proportion of the total population that identifies as non-Hispanic Black or African American alone in census tract $t$

$X_t^{asian}$ is the proportion of the total population that identifies as non-Hispanic Asian alone in census tract $t$

$X_t^{amerind}$ is the proportion of the total population that identifies as American Indian and Alaska native alone in census tract $t$

$X_t^{haw}$ is the proportion of the total population that identifies as Hawaiian and other Pacific Islanders alone in census tract $t$

$X_t^{other}$ is the proportion of the total population that identifies as some other race alone in census tract $t$

$X_t^{twomore}$ is the proportion of the total population that identifies as two or more races alone in census tract $t$

$X_t^{medinc}$ is the median income of the house hold in census tract $t$

$X_t^{totpop}$ is the total population in census tract $t$

$X_t^{area}$ is the area of the census tract $t$

To assess the factors that may affect whether the county is a net importer or exporter, we run a logit model specification at the county level. It is expressed as:

**Equation 3.8**

$$logit(p_c(x)) = log\left(\frac{p(x)}{1-p(x)}\right) = \eta_{p,c}(x)$$

$$\eta_{p,t}(x) = \beta_0 + \beta_{black}X_c^{black} + \beta_{asian}X_c^{asian} + \beta_{hisp}X_c^{hisp} + \beta_{amerind}log(X_c^{amerind}) + \beta_{haw}X_c^{haw}$$
$$+ \beta_{other}X_c^{other} + \beta_{twomore}X_c^{twomore} + \beta_{medinc}log(X_c^{medinc}) + \beta_{totpop}log(X_c^{totpop})$$
$$+ \beta_{area}log(X_c^{area}) + \epsilon_{p,c}$$

Where,

$p_c(x)$ is the probability of the county $c$ being a net importer of aggregated air pollution related human health damage for pollutants (PM$_{2.5}$, SO$_2$, and NO$_x$)

$X_c^{black}$ is the proportion of the total population that identifies as non-Hispanic Black or African American alone in county $c$

$X_c^{asian}$ is the proportion of the total population that identifies as non-Hispanic Asian alone in county $c$

$X_c^{amerind}$ is the proportion of the total population that identifies as American Indian and Alaska native alone in county $c$

$X_c^{haw}$ is the proportion of the total population that identifies as Hawaiian and other Pacific Islanders alone in county $c$

$X_c^{other}$ is the proportion of the total population that identifies as some other race alone in county $c$

$X_c^{twomore}$ is the proportion of the total population that identifies as two or more races alone in county $c$

$X_c^{medinc}$ is the median income of the house hold in county $c$

$X_c^{totpop}$ is the total population in county $c$

$X_c^{area}$ is the area of the county $c$

Supplementary Information for

# Environmental injustice in America: Racial disparities in exposure to air pollution health damages from freight trucking

**Priyank Lathwal,[a] Parth Vaishnav,[b,*] M. Granger Morgan[c]**

[a] Belfer Center for Science and International Affairs, Harvard University, Cambridge, MA 02138
[b] School for Environment and Sustainability, University of Michigan, Ann Arbor, MI 48109
[c] Department of Engineering and Public Policy, Carnegie Mellon University, Pittsburgh, PA 15213
*Corresponding Author; Email: parthtv@umich.edu

This PDF file includes:
   **Supplementary Notes**
   **Figures S1 to S5**
   **Tables S1 to S9**
   **References**

**Supplementary Notes**

**1. Freight trucking emissions in National Emissions Inventory (NEI), 2017**

Short haul and long-haul heavy-duty trucks account for ~81% of freight transportation's total medium and heavy-duty petroleum consumption by mode in the US (see Table 1.17) [2]. To estimate percentage of freight trucking's contribution to total air pollutant and greenhouse gas (GHG) emissions in the United States, we consider emissions from the following sources for the country included in the most recent NEI:

1. point
2. non-point
3. on-road
4. non-road
5. wildfire events

We only account for emissions in the contiguous US states and exclude Alaska, Hawaii, Puerto Rico, Virgin Islands, American Samoa, Guam, and other non-contiguous territories. Next, we merge the NEI with the source classification codes (SCCs) [1] that the US EPA utilizes to classify different activities that contribute to emissions. SCC values provide "a unique source category-specific process or function that emits air pollutants"[1]. We assume that a dominant share of US freight tonnage is carried by diesel fueled freight trucks. So, we filter the NEI 2017 for only those on-road vehicles where the fuel use was diesel. We observe that within the "on-road" source category, there are 16 vehicle categories where the fuel used is indicated as diesel fuel. These are

1. passenger truck
2. light commercial truck
3. single unit short-haul truck
4. single unit long-haul truck
5. refuse truck
6. combination short-haul truck
7. combination long-haul truck
8. truck
9. tank cars and trucks
10. automobiles/truck assembly operations
11. automobiles and light trucks
12. tank truck cleaning
13. intercity bus
14. transit bus
15. school bus
16. motor home

Out of these, we only include five truck categories (i.e. (3) single unit short-haul truck, (4) single unit long-haul truck, (5) refuse truck, (6) combination short-haul truck, (7) combination long-haul truck) that are relevant for heavy-duty freight trucking. We exclude all other vehicle categories because they are either passenger vehicle fleets or involved in local operations.

*Freight trucking emission estimates comparison:* As a check, we compare our derived emission estimates with emissions attributable to freight trucking in the NEI, 2017 [3]. We find that the estimated emissions diverge from values estimated by 2017 NEI between 11% and 47% for different pollutants and the directionality of the difference is not preserved across pollutant type. The largest differences arise for $PM_{2.5}$, where our emission estimate is smaller by 46% and $CO_2$ where our emission estimate is larger by 47%. The differences arise presumably because the base year of trucking activity data used in developing FAF4 is from HPMS that we project to 2017 using VMT projection guidance by the FHWA. In comparison, US EPA's NEI 2017 utilizes more recent data. **Table S1** provides freight trucking emissions contribution to total US air pollutants and GHG derived from NEI, 2017.

**Table S1. Freight trucking emissions obtained from NEI, 2017 data [3]. The difference in estimates column provides the percentage difference of diesel truck emission estimates derived from Highway Performance Monitoring System (HPMS) link-level activity data (included in FAF4, 2017) and US EPA's 2017 NEI diesel truck emissions. The estimates in the last two columns provide percentage contribution of freight trucking emissions to total US emissions from FAF4 and NEI data. These are obtained by dividing the FAF4 and NEI freight trucking emissions by emissions from all sources in NEI, 2017.**

| Pollutant/ GHG | Trucking emissions derived from HPMS data (tons) | Trucking emissions derived from NEI data (tons) | % Difference | Total NEI emissions (tons) | % US emissions from freight trucking based on HPMS data | % US emissions from freight trucking based on NEI 2017 data |
|---|---|---|---|---|---|---|
| $PM_{2.5}$ | 28K | 50K | -46% | 5.2M | 0.53% | 1.0% |
| $SO_2$ | 4.6K | 3.7K | 24% | 2.5M | 0.18% | 0.10% |
| $NO_x$ | 1.1M | 1.3M | -11% | 11M | 10% | 12% |
| $CO_2$ | 640M | 430M | 47% | 5.3B | 12% | 8.2% |

## 2. Road Network Data

FAF4 data consists of ~446,000 miles of roads consisting of interstate highways, urban and rural principal arterials [4]. It builds on the CFS data [5], which is a publicly available dataset and provides information on national freight flows in the US. FAF4 information is more comprehensive in coverage than the CFS data because it includes industries and shipments that are not included in CFS while providing shipment information at the county level [6]. Although a more recent version of the Freight Analysis Framework Version 5 (FAF5) is available, it has yet to be updated with information on county level freight activity. For truck counts, FAF4 data includes link-level trucking activity that are based on HPMS elements. For the analysis, we assume that long-haul highway trips are conducted by heavy duty tractor-trailer diesel trucks (class 8b or above) whereas all non-long haul highway freight trips are conducted by single unit trucks (class 6 trucks). Further, we estimate the road-length by using the difference between "ENDMP" and "BEGMP" to calculate the length of each given road link. HPMS data is compiled by the Federal Highway Administration based on actual truck activity reported by states. In our data, we had to drop 94 out of 670,045 road links for which the result was a negative road length, which we assumed to be erroneous or incomplete. This maybe because some of the field information might not be available at the time of reporting, resulting in missing values or incomplete information. Links with negative lengths account for ~0.01% of the total and dropping them should have a negligible effect on the results. Further, we compare our road link lengths against the "MILES" variable available in the data which was another variable that gave the length of the road link in miles. The cumulative distribution function (CDF) in **Figure S1** indicates that both these variables are nearly identical. We multiply the reported truck counts for each road link with length of the road link. This gives us the vehicle miles traveled (VMT).

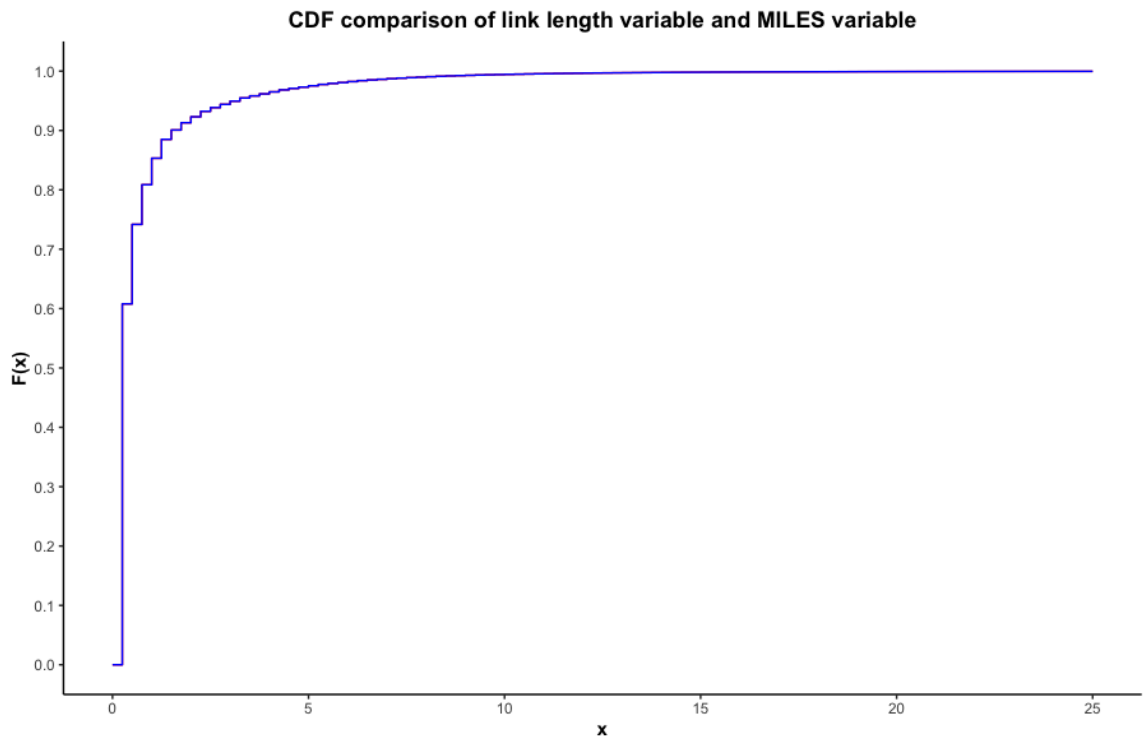

**Figure S1.** CDF comparing our road length estimate with the "MILES" variable included in the FAF4 data. These are almost similar. So, the impact of dropping some of the inaccurate road links has negligible impact on the results.

### 3. Emission Factors for Single Unit and Combination Trucks

Even though there may be variation in emission factors for trucks on different types of road, we use emission factor for single unit and combination trucks from the Greenhouse gases, Regulated Emissions, and Energy use in Technologies (GREET) model [7]. Further, we also account for tire and brake wear emissions. **Table S2** reproduces the emission factors for $PM_{2.5}$, $SO_2$, $NO_x$, and $CO_2$ used in the study.

**Table S2. Lifetime mileage weighted emission factors (in g/mile) for single unit and combination trailer freight diesel trucks. These values have been taken from the GREET model [7].**

| Pollutant/ GHG | Emission factor for combination trailer truck (in g/mile) | Emission factor for single unit truck (in g/mile) |
|---|---|---|
| $PM_{2.5}$ | 0.086 | 0.0467 |
| $SO_2$ | 0.0149 | 0.0070 |
| $NO_x$ | 4.585 | 0.9383 |
| $CO_2$ | 1588 | 1414 |

As a sensitivity check, we also compare our long-haul trucking results by running the analysis using emission factors from Table S8 in Tong et al.[8]. Using an average fuel economy of 6.3 miles per diesel gallon equivalent [9] for long-haul freight trucking, we convert the emission factors to g/mile. Next, we weight the emission factors using lifetime miles for a combination tractor (Table 2-28)[10] to estimate lifetime mileage weighted emission factors. Tong et al.'s [8] method is different than ours since the authors rely on a mass-balance approach and harmonize emissions from literature for different air pollutants and greenhouse gases. **Table S3** compares emissions for long-haul freight trucking using modified emission factors from Tong et al.[8] and our emissions based on emission factors from the GREET model [7].

**Table S3. Comparison of long-haul freight trucking emissions (in tons) using lifetime mileage weighted emission factors reported in Tong et al. [8] and GREET model [7].**

| Pollutant/ GHG | Long-haul trucking emissions (in tons) using Tong et al. emission factors | Long-haul trucking emissions (in tons) using GREET emission factors |
|---|---|---|
| $PM_{2.5}$ | 5.5K | 17K |
| $SO_2$ | 80 | 3K |
| $NO_x$ | 108K | 920K |
| $CO_2$ | 32M | 31M |

## 4. Reduced-Complexity Chemical Transport Models (CTMs)

Social costs arising from exposure to fine particulate matter ($PM_{2.5}$) on human health are critical for designing effective air pollution control policies. However, the tools available to link changes in $PM_{2.5}$ emissions and subsequent impacts on public health effects are limited. Usually, such situations require deploying and running a state-of-the-science CTM, however, the process can be computationally expensive and time consuming. To overcome these limitations and facilitate a quick estimation of social costs on human health due to air pollution, researchers have developed reduced-complexity models (RCMs) that provide tabulated marginal social costs for different pollutant species. In this study, we employ RCMs to estimate air pollution related public health damages from freight trucking. The subsequent section will discuss the models employed.

### 4.1. Estimating Air Pollution Social Impacts Using Regression (EASIUR)

In this study, we estimate the impacts of freight trucking pollution on human health at the county level using an RCM called EASIUR [11,12]. The model provides estimates of marginal social costs for four species— (1) primary fine particulate matter ($PM_{2.5}$) species (elemental carbon), and three secondary inorganic species, namely, sulfur dioxide ($SO_2$), oxides of ($NO_x$), and ammonia ($NH_3$). These inorganic species are responsible for secondary formation of $PM_{2.5}$. These estimates of marginal social costs were derived by running regressions on the results of a CTM. The EASIUR model maps the contiguous United States in 148 x 112 grid cells where each grid cell is 36 km x 36 km and is available for four seasons and for three emission elevations: ground level and two stack heights (150 m and 300 m) for point sources. For the scope of this analysis, we ignore any seasonal variation in marginal social costs arising from freight trucking pollution and only consider annual impacts of freight trucking on public health at the ground level. Additionally, the model also comes along with Air Pollution Social Cost Accounting (APSCA) model (available here: https://barney.ce.cmu.edu/~jinhyok/apsca/) that allows for estimating source-receptor (S-R) relationships for marginal changes in emissions for different pollutant species at a fine spatial resolution [12].

### 4.2. Comparison of freight trucking air pollution related public health social costs from different models

As a check, we compare our freight trucking public health damage results by running the analysis on two emissions inventories: (1) FAF4 and (2) NEI, 2017. Additionally, we compare the results using another model called the Air Pollution Emission Experiments and Policy Version 3 (AP3) [*][13] on both the emission inventories. This is an updated version of the Air Pollution Emission Experiments and Policy Version 2 (AP2) model that employs (S-R) matrices to map ambient concentrations to respective counties. The contributions are reflected as an individual element in the S-R matrix. The model is available for four heights: (1) ground-level, (2) point sources located less than 250 m in height, (3) point sources located between 250 m to 500 m in height, and (4) point sources greater than 500 m in height. We use only the ground-level sources to assess the annual impacts of freight trucking on public health.

To estimate freight trucking public health damages, first, we require the EASIUR marginal damages for 2017. To do this, we use the approach suggested by Heo et al. [11] in the supplementary information. The EASIUR marginal damage estimates can be adjusted for a VSL for 2017 ($VSL_{2017}$) using the following adjustment factor ($F_{2017}$):

$$F_{2017} = \frac{VSL_{2017}}{VSL_{2010}}$$

Where,
$F_{2017}$ is the adjustment factor by which the EASIUR marginal damages are adjusted
$VSL_{2017}$ is the VSL value for the year 2017
$VSL_{2010}$ is the VSL value for the year 2010 and it's value is \$8.6M

---

[*]The marginal social costs for pollutant species for the updated AP3 model was obtained from the authors of Tschofen et al. [17]

Next, we need to estimate $VSL_{2017}$. At the time the EASIUR model was released, it used a VSL value of $8.6M while relying on population year and income year 2005. We update the population year to 2017 for estimating public health damages by modifying the EASIUR python script. Additionally, we also need to consider income growth and inflation. We do this by using income growth adjustment factors that come along with the EASIUR model. First, we calculate the updated VSL for the income year 2017 using the following equation:

$$VSL_{2017} = VSL_{2010} * \left(\frac{I_{2017}}{I_{2010}}\right) * \left(\frac{CPI_{2017}}{CPI_{2010}}\right)$$

$$VSL_{2017} = \$8.6M * \left(\frac{1.174}{1.010}\right) * \left(\frac{245}{218}\right)$$

$$VSL_{2017} = \$10.3M$$

Where,
$VSL_{2017}$ is the VSL value for the year 2017
$VSL_{2010}$ is the VSL value for the year 2010
$I_{2017}$ is the income growth adjustment factor in the EASIUR model and its value is 1.174
$I_{2010}$ is the income growth adjustment factor in the EASIUR model and its value is 1.010
$CPI_{2017}$ is the consumer price index (CPI) for the year 2017 and is obtained from the U.S. Bureau of Labor Statistics (BLS) [14]
$CPI_{2010}$ is the CPI for the year 2010 and is obtained from the U.S. (BLS) [14]

This calculation gives us a $VSL_{2017}$ of $10.3M in 2017. The VSL report by the US Department of Transportation (DOT) in 2017 was $10.2M [15].

### 4.2.1. Public health damage results based on the emissions derived from FAF4 data

**Table S4** provides absolute values of air pollution related public health damages from freight trucking in 2017 US $ based on emissions derived from FAF4.

**Table S4. Premature mortality and public health damages (in 2017 US $) due to freight trucking for EASIUR and AP3 model. These results are based on the bottom-up emissions inventory derived from FAF4 and the VSL value used is $10.3M in 2017 US $.**

| Model | PM$_{2.5}$ | | SO$_2$ | | NO$_x$ | |
|---|---|---|---|---|---|---|
| | Premature mortality | Trucking damages (in US $) | Premature mortality | Trucking damages (in US $) | Premature mortality | Trucking damages (in US $) |
| EASIUR | 527 | 5.5B | 11 | 110M | 1,076 | 11B |
| AP3 | 552 | 5.7B | 19 | 370M | 2,381 | 25B |

For the emissions derived from HPMS data included in FAF4, we also compare county level results across EASIUR and AP3. The social costs are broadly consistent with a slight deviation from the y-x line for SO$_2$ and NO$_x$ damages (see **Figure S2**). This is because of other associated uncertainties and in general, the cost estimates for secondary pollutants are found to be more variable and the correspondence in terms of Pearson's correlation for secondary pollutants marginal damages, especially for NO$_x$ and SO$_2$ damages across both models is lower [16].

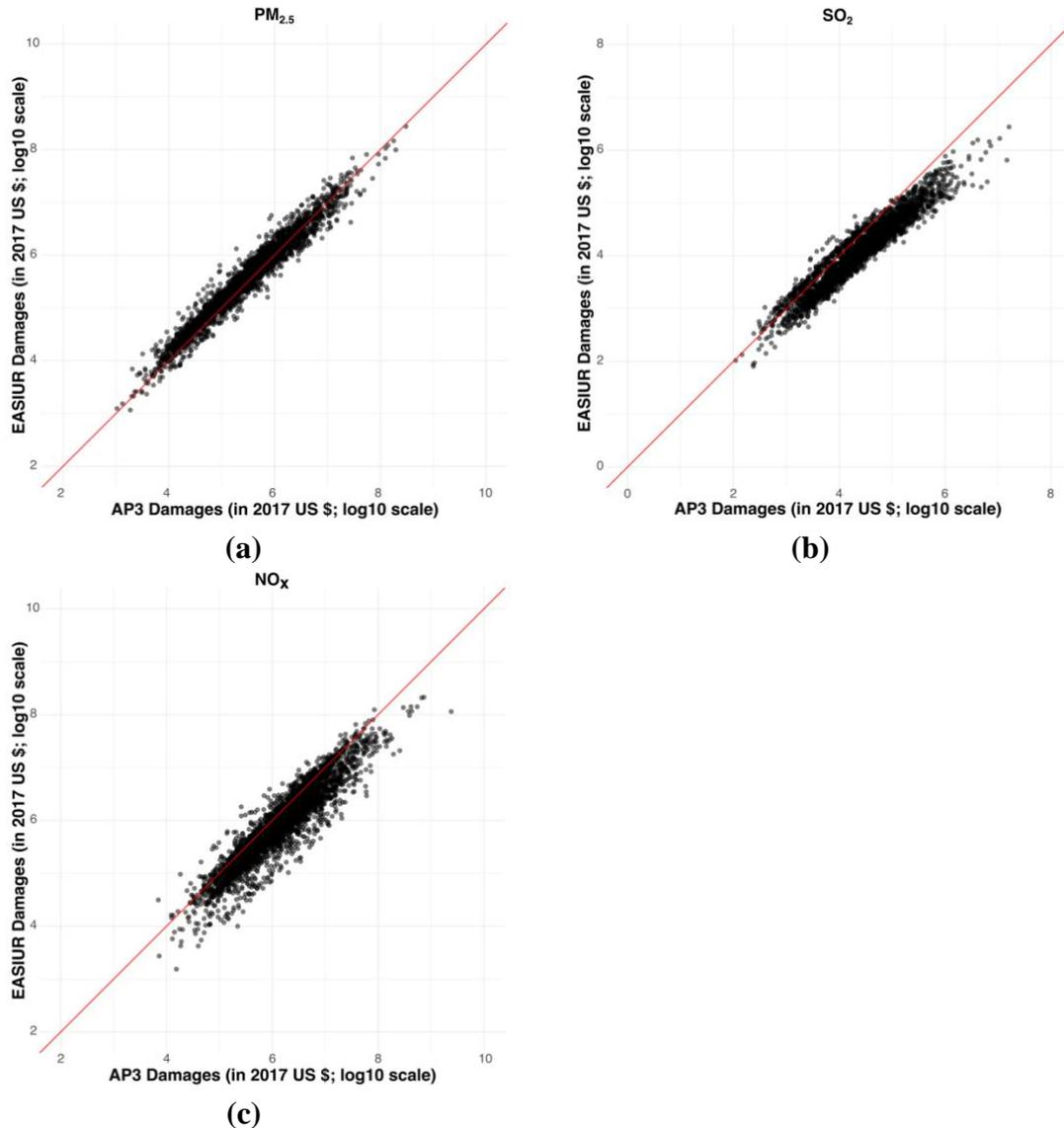

**Figure S2. Comparison of freight trucking air pollution related public health social costs (in log10 tons) from EASIUR and AP3. These damages are based on emissions derived from FAF4 data. We observe that social costs from both the models roughly lie on the y=x line.**

### 4.2.2. Public health damage results based on the NEI emissions inventory

**Table S5** provides absolute values of air pollution related public health damages from freight trucking in 2017 US $ based on the NEI 2017 emissions inventory. As in the case of previous comparison, here, we compare county level results across EASIUR and AP3 for the NEI 2017 emissions inventory. The social costs are broadly consistent with a slight deviation from the y-x line for SO$_2$ and NO$_x$ damages (see **Figure S3**).

**Table S5. Premature mortality and public health damages (in 2017 US $) due to freight trucking for EASIUR and AP3 model. These results are based on the NEI emissions inventory compiled by the authors and the VSL value used is $10.3M in 2017 US $.**

| Model | PM$_{2.5}$ | | SO$_2$ | | NO$_x$ | |
|---|---|---|---|---|---|---|
| | Premature mortality | Trucking damages (in US $) | Premature mortality | Trucking damages (in US $) | Premature mortality | Trucking damages (in US $) |
| EASIUR | 1,160 | 12B | 10 | 100M | 1,552 | 16B |
| AP3 | 1,270 | 13B | 37 | 384M | 3,684 | 38B |

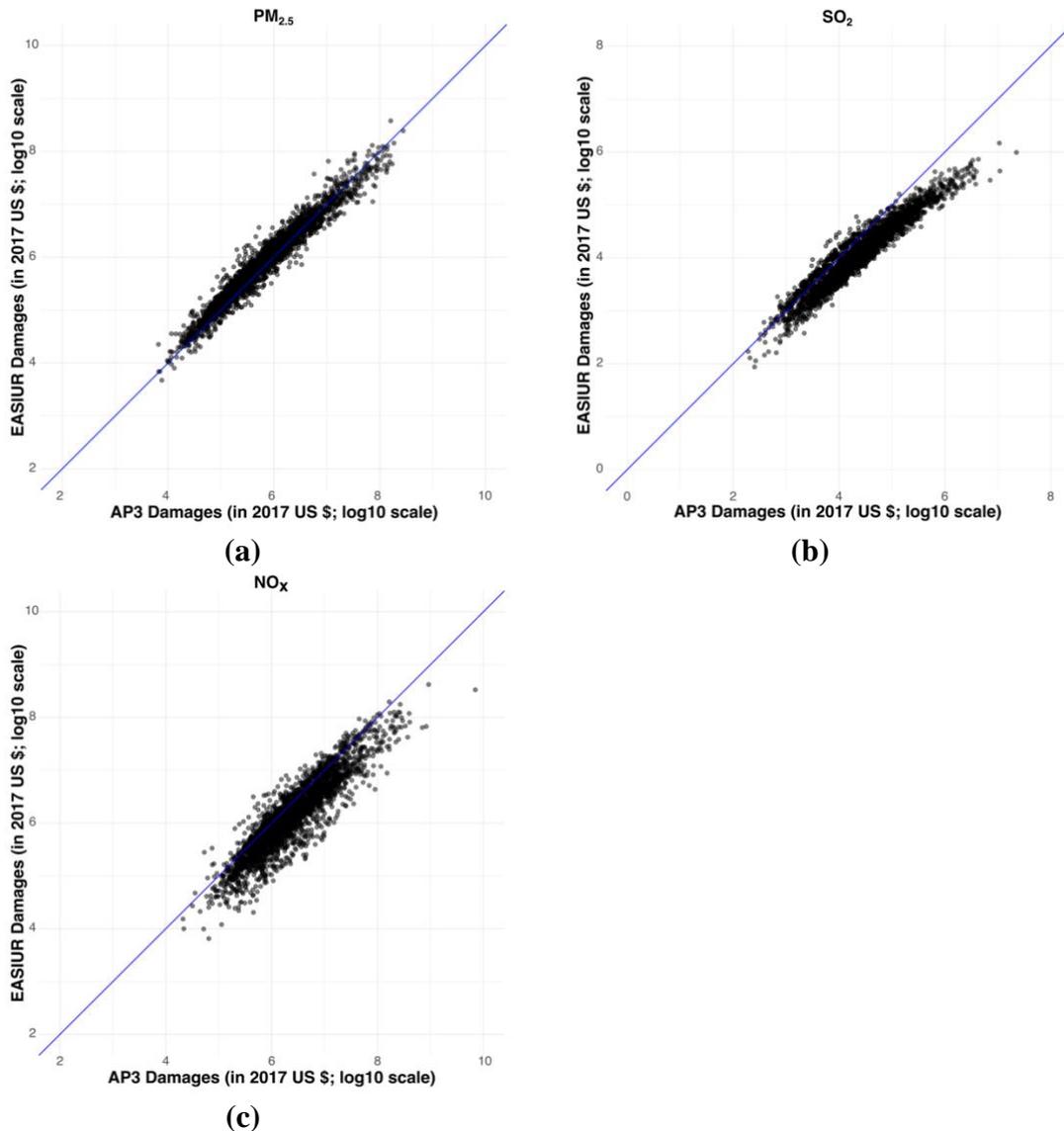

**(a)**

**(b)**

**(c)**

**Figure S3. Comparison of freight trucking air pollution related public health social costs (in log10 tons) from EASIUR and AP3. These damages are based on the NEI emissions inventory. We observe that social costs from both the models roughly lie on the y=x line.**

# 5. Distributional effects of freight trucking air pollution at the county level

We run a demographic analysis to evaluate the demographic impacts of freight trucking air pollution at the county level on different racial-ethnic subgroups. We provide a description of the dependent and independent variables used in **Table S6** along with the hypothesized relationship. Next, we apportion the freight trucking emissions in respective census tracts and repeat the analysis at the census tract level using the same demographic variables. We run three model specifications for $PM_{2.5}$, $SO_2$, and $NO_x$ emissions from freight trucking at the county and census tract level.

**Table S6. Description of variables used in the regression analysis at the county level. We also conduct the analysis using the same variables at the census tract level.**

| Variable Type | Definition | Acronym | Units | Hypothesized Relationship |
|---|---|---|---|---|
| Dependent | Freight trucking emissions for pollutant $p$, where is $PM_{2.5}$, $SO_2$, and $NO_x$ in county $c$ | $Y_{p,c}$ | Emissions in tons | |
| Independent | proportion of the total population in the county $c$ that identifies as non-Hispanic Black or African American | $X_c^{black}$ | Proportion of people | Positive |
| Independent | proportion of the total population in the county $c$ that identifies as non-Hispanic Asian alone | $X_c^{asian}$ | Proportion of people | Positive |
| Independent | proportion of the total population in the county $c$ that identifies as American Indian and Alaska native alone | $X_c^{amerind}$ | Proportion of people | Positive |
| Independent | proportion of the total population in the county $c$ that identifies as Hawaiian and other Pacific Islanders alone | $X_c^{haw}$ | Proportion of people | Positive |
| Independent | proportion of the total population in the county $c$ that identifies as some other race alone | $X_c^{other}$ | Proportion of people | Positive |
| Independent | is the proportion of the total population in county $c$ that identifies as two or more races alone | $X_c^{twomore}$ | Proportion of people | Positive |

| Independent | median household income in county $c$ | $X_c^{medinc}$ | In 2017 US dollars | Negative |
| --- | --- | --- | --- | --- |
| Independent | total population in county $c$ | $X_c^{totpop}$ | Number of people | Positive |
| Independent | area of the county $c$ | $X_c^{area}$ | Acres | Positive |

### 5.1 Distribution of dependent variables in FAF4 data

We hypothesize that the errors in our regression relationship function $y_i = \beta_0 + \beta_i x_i + \epsilon_i$ are normally distributed such that $\epsilon_i \sim N(0, \sigma^2)$. Next, we look at plot the quantiles of our dependent variables against quantiles of normal distribution. The advantage of using a Q-Q plot is that it allows us to simulate as many draws from the normal distribution as possible to satisfactorily represent the distribution. **Figure S4** shows Q-Q plots for dependent variables. We see that the log-transformed variables are much closer to the normal distribution.

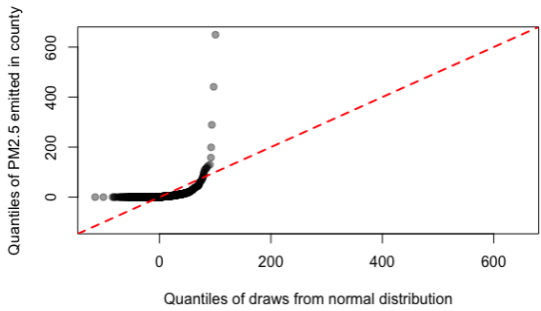

**(a)**

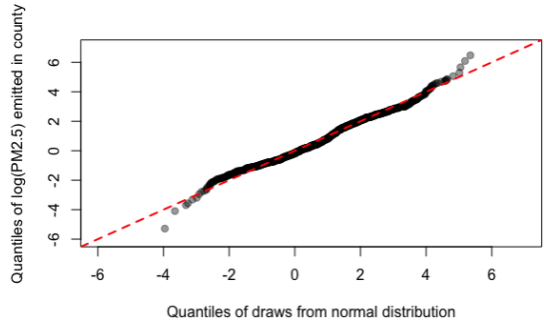

**(b)**

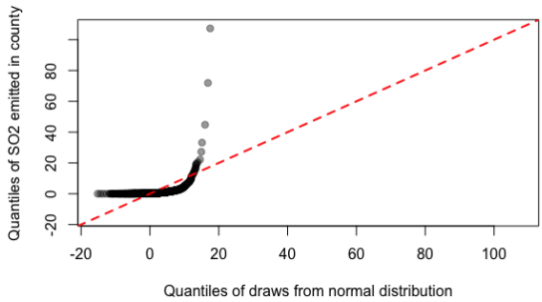

**(c)**

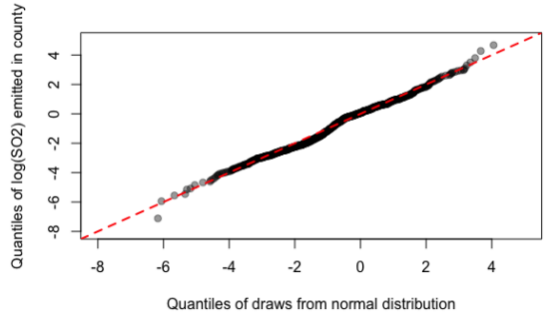

**(d)**

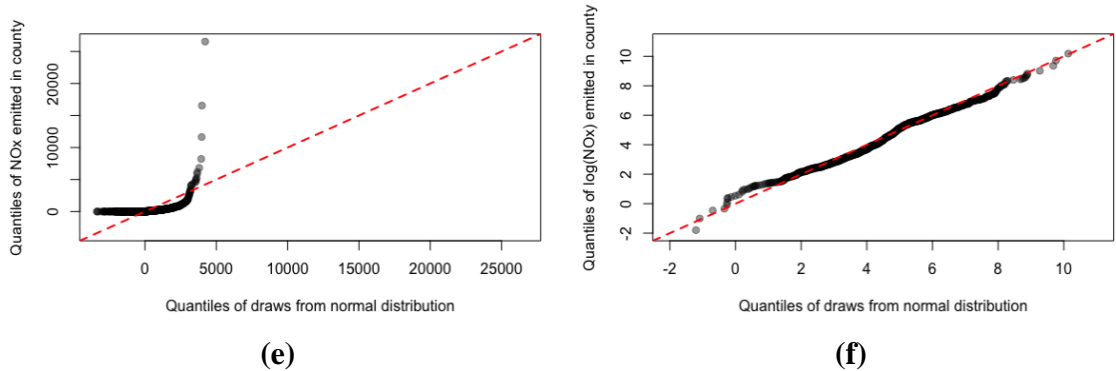

<div align="center">(e)             (f)</div>

**Figure S4. Q-Q plot of untransformed dependent variables (PM$_{2.5}$, SO$_2$, and NO$_x$ emissions). The untransformed variables are non-linear (a,c,e; left panel) whereas the log transformed dependent variables, i.e., log(PM$_{2.5}$), log(SO$_2$), and log (NO$_x$) emissions (b,d,f; right panel) are distributed normally.**

### 5.2 Distribution of independent variables

We calculate proportions of different racial and ethnic sub-groups at the county level from the total population provided in the census data. This is done in a manner that these groupings are mutually exclusive. Additionally, we log transform the area of the county, median county level household income, and the total population of the county. **Figure S5**. shows the histograms of county relevant independent variables included in the specification. We repeat the analysis for the same variables at the census tract level and the results are enclosed in subsequent sections.

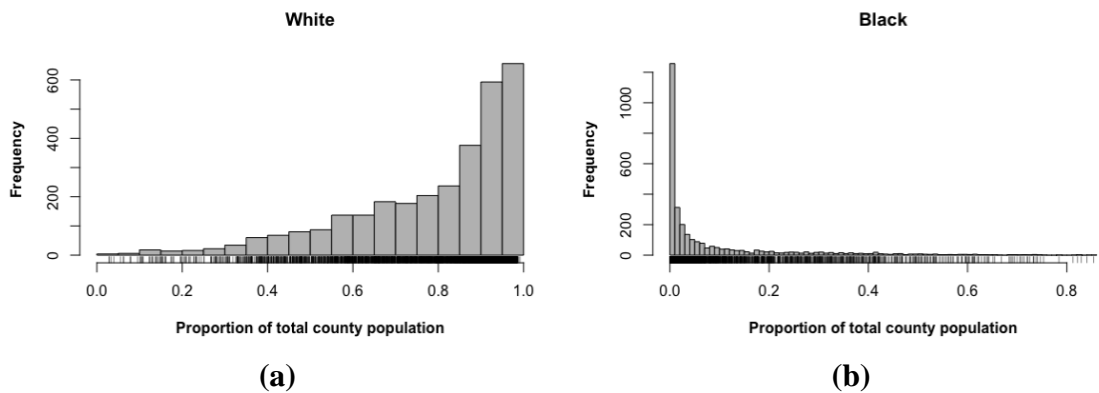

<div align="center">(a)             (b)</div>

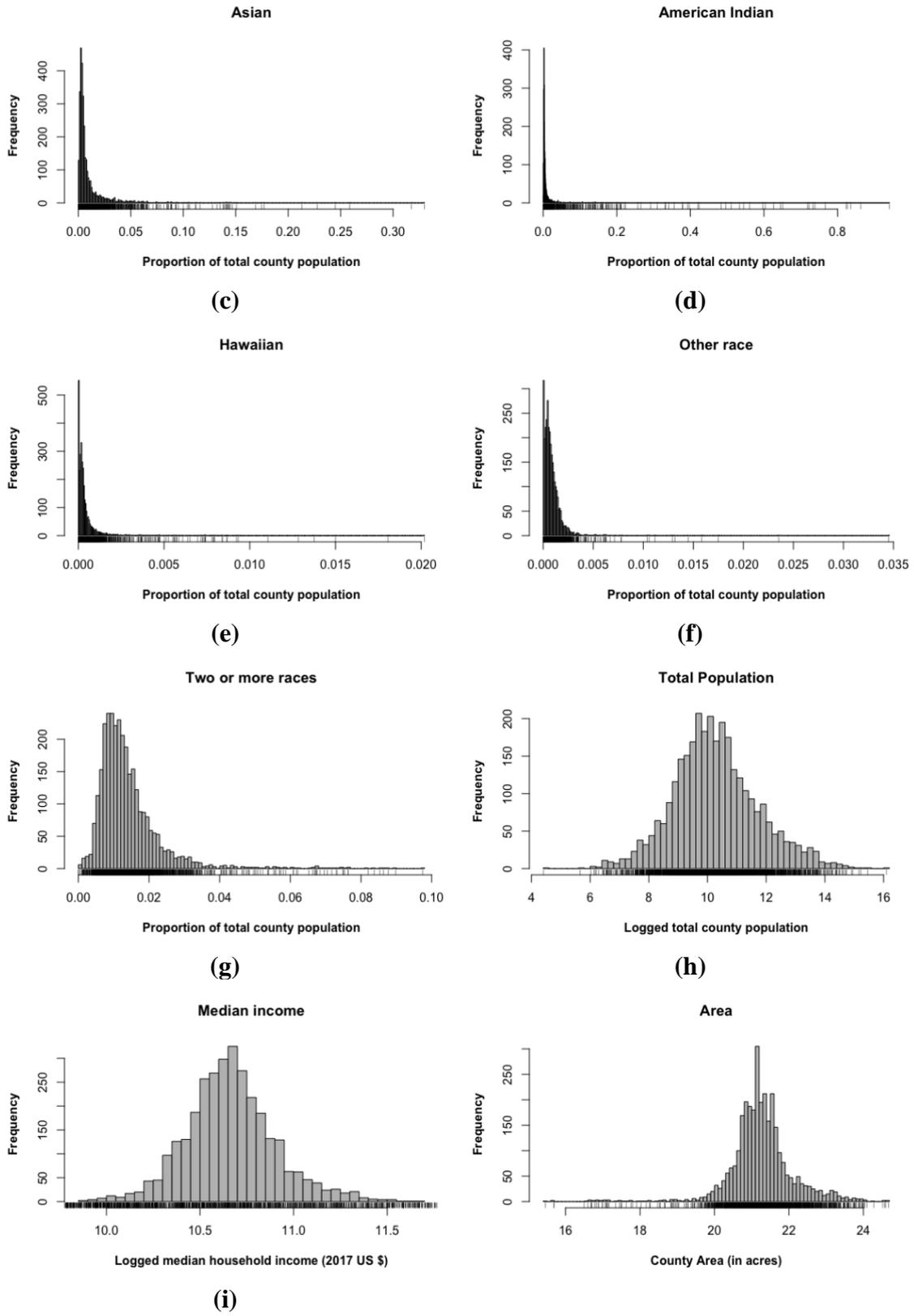

**Figure S5. Distribution of independent variables that are included in the model specification.**

### 5.3 Model specification results due to freight trucking pollution at the county and census tract level

In this section, we compare distributional impacts of air pollution from freight trucks at the county level. We run regression specifications on freight trucking emissions derived from the NEI 2017 emissions inventory that is is publicly available from the US EPA website [3]. **Table S7** and **Table S8** show the results of running the models at the county and census tract level for $PM_{2.5}$, $SO_2$, and $NO_x$ emissions. For the census tract results, we down scale the county level emission estimates from NEI 2017 to the census tract by population weighting approach. We observe higher levels of trucking air pollution on disadvantaged communities (DACs) in the NEI 2017 emission inventory at the county level. The results are similar to what we observe in emissions derived from the FAF4 data. Counties with higher proportions of Black and Hispanic residents are more exposed to higher emissions from freight trucks.

**Table S7. Comparison of demographic effects of $PM_{2.5}$, $SO_2$, and $NO_x$ freight trucking emissions at the county level. These results are based on emissions from freight trucking included in the NEI 2017 emissions inventory. The numbers in the parenthesis provide standard errors.**

|  | *Dependent variable:* | | |
|---|---|---|---|
|  | log(PM2.5) | log(SO2) | log(NOx) |
|  | (1) | (2) | (3) |
| Black | 0.650*** | 0.889*** | 0.601*** |
|  | (0.088) | (0.098) | (0.095) |
| Asian | -3.421*** | -4.576*** | -2.992*** |
|  | (0.726) | (0.809) | (0.781) |
| Hispanic | 0.345*** | 0.885*** | 0.642*** |
|  | (0.093) | (0.104) | (0.100) |
| American Indian/Alaskan Native | -0.262 | -0.535** | -0.291 |
|  | (0.202) | (0.225) | (0.217) |
| Native Hawaiian/ Other Pacific Islander | 0.490 | 1.473 | 3.152 |
|  | (11.238) | (12.519) | (12.074) |
| Other Race | -19.289* | -41.848*** | -30.889*** |
|  | (10.130) | (11.285) | (10.884) |
| Two or More Races | 2.914** | 2.867* | 2.890* |
|  | (1.431) | (1.594) | (1.538) |
| log(Median Income) | 0.252*** | 0.408*** | 0.317*** |
|  | (0.059) | (0.065) | (0.063) |
| log(Total Population) | 0.716*** | 0.733*** | 0.694*** |
|  | (0.010) | (0.011) | (0.011) |
| log(County Area) | 0.317*** | 0.320*** | 0.360*** |
|  | (0.015) | (0.016) | (0.016) |

| | | |
|---|---|---|
| Constant | -14.853*** | -19.505*** | -13.054*** |
| | (0.677) | (0.755) | (0.728) |
| Observations | 3,106 | 3,106 | 3,106 |

*Note:* *p<0.1; **p<0.05; ***p<0.01

**Table S8. Comparison of demographic effects of PM$_{2.5}$, SO$_2$, and NO$_x$ freight trucking emissions at the census tract level. These results are based on emissions from freight trucking included in the NEI 2017 emissions inventory. The numbers in the parenthesis provide standard errors.**

| | *Dependent variable:* | | |
|---|---|---|---|
| | log(PM2.5) | log(SO2) | log(NOx) |
| | (1) | (2) | (3) |
| Black | 1.003*** | 1.050*** | 0.953*** |
| | (0.020) | (0.022) | (0.021) |
| Asian | 0.730*** | 0.745*** | 1.057*** |
| | (0.054) | (0.057) | (0.055) |
| Hispanic | 1.522*** | 1.763*** | 1.911*** |
| | (0.022) | (0.023) | (0.022) |
| American Indian/Alaskan Native | 1.253*** | 0.977*** | 1.157*** |
| | (0.090) | (0.095) | (0.093) |
| Native Hawaiian/ Other Pacific Islander | 7.432*** | 8.600*** | 11.113*** |
| | (1.172) | (1.240) | (1.211) |
| Other Race | -9.481*** | -13.130*** | -14.482*** |
| | (0.768) | (0.813) | (0.794) |
| Two or More Races | 1.202*** | 2.027*** | 3.798*** |
| | (0.360) | (0.381) | (0.372) |
| log(Median Income) | 0.681*** | 0.709*** | 0.703*** |
| | (0.010) | (0.010) | (0.010) |
| log(Total Population) | 0.219*** | 0.236*** | 0.203*** |
| | (0.008) | (0.009) | (0.009) |
| log(County Area) | -0.308*** | -0.300*** | -0.284*** |
| | (0.002) | (0.002) | (0.002) |
| Constant | -0.942*** | -4.213*** | 1.668*** |
| | (0.122) | (0.129) | (0.126) |
| Observations | 71,038 | 71,038 | 71,038 |

*Note:* *p<0.1; **p<0.05; ***p<0.01

>15

We also assess whether a county is more likely to be an importer of air pollution or not. To evaluate this, we run logistic regression specifications at the county level. **Table S9** shows the results of logistic regression at the county level.

**Table S9. Logistic regression results of a county being an importer. These results are based on the NEI 2017 and trucking emissions derived from FAF4. The numbers in the parenthesis provide standard errors. Increasing the predictor by 1-unit results in multiplying the odds of having the outcome by $e^\beta$.**

|  | *NEI 2017* | *FAF4* |
|---|---|---|
|  | Importer County | Importer County |
| Black | 5.426*** | 0.268 |
|  | (0.456) | (0.297) |
| Asian | -21.849*** | 0.091 |
|  | (5.593) | (2.750) |
| Hispanic | 3.824*** | -1.504*** |
|  | (0.518) | (0.337) |
| American Indian/Alaskan Native | -1.915** | -2.109*** |
|  | (0.813) | (0.787) |
| Native Hawaiian/ Other Pacific Islander | -63.291 | -30.148 |
|  | (46.229) | (38.708) |
| Other Race | 91.597** | 103.442** |
|  | (40.273) | (52.737) |
| Two or More Races | 24.375*** | 23.437*** |
|  | (5.947) | (5.245) |
| log(Median Income) | -3.900*** | -0.747*** |
|  | (0.269) | (0.200) |
| log(Total Population) | -0.405*** | 0.344*** |
|  | (0.043) | (0.035) |
| log(County Area) | 1.068*** | -0.116** |
|  | (0.072) | (0.050) |
| Constant | 22.246*** | 6.811*** |
|  | (2.912) | (2.315) |
|  |  |  |
| Observations | 3,104 | 3,107 |